\documentclass{ws-procs9x6}
\usepackage{bbm}
\usepackage{graphics}
\usepackage{amsmath}
\usepackage{amsfonts,amsbsy}
\usepackage{amssymb}

\def\empile#1\over#2{\mathrel{\mathop{\kern 0pt#1}\limits_{#2}}}

\def\wt#1{\widetilde{#1}}

\newcommand{\slv}{\raise.15ex\hbox{$/$}\kern-.53em\hbox{$v$}}
\newcommand{\slF}{\raise.15ex\hbox{$/$}\kern-.53em\hbox{$F$}}
\newcommand{\slL}{\raise.15ex\hbox{$/$}\kern-.53em\hbox{$L$}}
\newcommand{\slP}{\raise.15ex\hbox{$/$}\kern-.53em\hbox{$P$}}
\newcommand{\slp}{\raise.15ex\hbox{$/$}\kern-.53em\hbox{$p$}}
\newcommand{\slq}{\raise.15ex\hbox{$/$}\kern-.53em\hbox{$q$}}
\newcommand{\slR}{\raise.15ex\hbox{$/$}\kern-.53em\hbox{$R$}}
\newcommand{\slQ}{\raise.15ex\hbox{$/$}\kern-.53em\hbox{$Q$}}
\newcommand{\slK}{\raise.15ex\hbox{$/$}\kern-.53em\hbox{$K$}}
\newcommand{\slk}{\raise.15ex\hbox{$/$}\kern-.53em\hbox{$k$}}
\newcommand{\slD}{\raise.15ex\hbox{$/$}\kern-.53em\hbox{$D$}}
\newcommand{\slC}{\raise.15ex\hbox{$/$}\kern-.53em\hbox{$C$}}
\newcommand{\slA}{\raise.15ex\hbox{$/$}\kern-.53em\hbox{$A$}}
\newcommand{\slSigma}{\raise.15ex\hbox{$/$}\kern-.53em\hbox{$\Sigma$}}
\newcommand{\slpartial}{\raise.15ex\hbox{$/$}\kern-.53em\hbox{$\partial$}}
\newcommand{\slcalP}{\raise.15ex\hbox{$/$}\kern-.63em\hbox{$\cal P$}}

\def\q{{\boldsymbol q}}
\def\l{{\boldsymbol l}}
\def\k{{\boldsymbol k}}

\def\x{{\boldsymbol x}}
\def\y{{\boldsymbol y}}

\begin{document}

\title{Proton-nucleus collisions\\ in the color glass condensate framework}

\author{J.-P.~Blaizot, F.~Gelis, R.~Venugopalan}

\address{CEA/DMS/SPhT\\
91191, Gif-sur-Yvette cedex, France}

\maketitle

\abstracts{We discuss proton-nucleus collisions in the framework of
the color glass condensate. By assuming that the proton can be
described as a low density color source, we solve exactly the
Yang-Mills equations corresponding to this type of collision, and then
use this solution in order to calculate inclusive gluon production or
quark-antiquark production. Our result shows that
$k_\perp$-factorization, while valid for gluon production, is violated
for quark pair production in proton-nucleus collisions.}

\section{Introduction}
When one increases the energy of a hadron (or equivalently when one
looks at partons with a smaller and smaller momentum fraction $x$),
its parton density increases. At moderate energies, this increase is
governed by the linear BFKL equation. This equation leads to a growth
of the gluon distribution $xG(x,Q^2)$ as a positive power of $1/x$
which would lead to violations of unitarity. However, it has been
known for quite some time that processes which are of higher order in
the gluon density -- like the recombination of two gluons -- become
important at small $x$. These non-linear effects in the evolution
limit the growth of the gluon density \cite{GLR-MQ}.

Note that these non-linear effects will arise sooner in a large
nucleus than in a proton, because its parton density per unit of
transverse area is enhanced by a factor $A^{1/3}$ where $A$ is the
atomic number of the nucleus. Therefore, there is a window in energy
for which the proton can be considered to a be a dilute projectile
while the nucleus is a dense projectile. In this talk, we address this
situation in the framework provided by the Color Glass
Condensate\cite{CGC}. In this description, the projectiles are
effectively described as static random sources of color charge on the
light-cone \cite{MV}. In order to evaluate a given process in this
framework, one should first calculate the classical color field
(i.e. solve the classical Yang-Mills equations) generated by these
sources of color, evaluate the matrix element of interest in the
presence of this classical field, and average over the random color
sources of both projectiles.

The distribution of the random sources $\rho$ of a projectile is some
functional $W_x[\rho]$ that depends on the typical momentum fraction
$x$ probed in the projectile by the process one considers. The Color
Glass Condensate predicts the evolution of this distribution when $x$
decreases, via the JIMWLK equation \cite{JIMWLK,CGC}. This equation is
a Wilson renormalization group equation that reduces to the BFKL
equation in the low density limit. Note however that the initial
condition for this evolution cannot in general be predicted from
perturbative QCD, although for a large nucleus it is expected that a
Gaussian distribution might be a good ansatz (as in the
McLerran-Venugopalan model \cite{MV}).

\section{Solution of the classical Yang-Mills equations}
The first step in using the color glass condensate framework in order
to calculate a given process is to solve the classical Yang-Mills
equations for two sources representing the colliding hadrons. This
cannot be done analytically in the general case, but it turns out that
an exact solution exists in the limit where the classical source
representing the proton is treated to first order while all orders in
the source representing the nucleus are kept. More precisely, one
needs to solve the equation $[D_\mu,F^{\mu\nu}]=J^\nu$ where, at
lowest order in the sources,
\begin{eqnarray}
J^\nu_a=
g\delta^{\nu+}\delta(x^-)\rho_{p,a}(\x_\perp)
+
g\delta^{\nu-}\delta(x^+)\rho_{_{A},a}(\x_\perp)\; .
\label{eq:sources}
\end{eqnarray}
Corrections to the current $J^\mu$ are determined from the requirement
that it must be covariantly conserved: $[D_\nu,J^\nu]=0$.  Finally, in
order to close the set of equations, one must also impose a gauge
condition to the field $A^\mu$. In this work, we use the Lorenz gauge,
$\partial_\mu A^\mu=0$.  Indeed, in this gauge, the solution has a
very simple expression which turns out to be very convenient when we
later calculate the pair production cross-section.

It is fairly straightforward to solve this set of equations to first
order in the source $\rho_p$ of the proton, and the Fourier transform
of the solution reads \cite{BGV1}:
\begin{eqnarray}
{ A}^\mu(q)&=
&\frac{ig}{q^2+iq^+\epsilon}
\int\frac{d^2\k_{1\perp}}{(2\pi)^2}
\Big\{
C_{_{U}}^\mu(q,\k_{1\perp})\, 
\big[U(\k_{2\perp})-(2\pi)^2\delta(\k_{2\perp})\big]
\nonumber\\
&&\qquad\qquad
+
C_{_{V}}^\mu(q)\, 
\big[V(\k_{2\perp})-(2\pi)^2\delta(\k_{2\perp})\big]
\Big\}
\frac{{ \rho}_p(\k_{1\perp})}{k_{1\perp}^2}
\; ,
\label{eq:A1infty-final}
\end{eqnarray}
where implicitly $\k_{2\perp}\equiv\q_\perp-\k_{1\perp}$. The 4-vectors $C_{_{U,V}}^\mu$ are,
\begin{eqnarray}
&& 
C_{_{U}}^+(q,\k_{1\perp})\equiv -\frac{k_{1\perp}^2}{q^-+i\epsilon}
\;,\;
C_{_{U}}^-(q,\k_{1\perp})\equiv \frac{k_{2\perp}^2-q_\perp^2}{q^+}
\;,\; 
C_{_{U}}^i(q,\k_{1\perp})\equiv -2 k_1^i\; ,
\nonumber\\
&& 
C_{_{V}}^+(q)\equiv 2q^+ \quad,\quad 
C_{_{V}}^-(q)\equiv 2\frac{q_\perp^2}{q^+}-2q^-\quad,\quad 
C_{_{V}}^i\equiv 2 q^i\; ,
\end{eqnarray}
and the quantities $U, V$ are Fourier transforms of Wilson lines:
\begin{eqnarray}
&&
U(\k_\perp)\equiv 
\int d^2\x_\perp e^{-i\k_\perp\cdot\x_\perp} 
{\mathcal P}e^{ig\int dz^+ A_{_A}^-(z^+,\x_\perp)\cdot T}
\; ,\nonumber\\
&&
V(\k_\perp)\equiv 
\int d^2\x_\perp e^{-i\k_\perp\cdot\x_\perp} 
{\mathcal P}e^{i\frac{g}{2}\int dz^+ A_{_A}^-(z^+,\x_\perp)\cdot T}
\; .
\end{eqnarray}
In these Wilson lines, the field $A_{_{A}}^\mu$ is the field generated
by the source $\rho_{_{A}}$ of the nucleus. All the higher-order
effects in the strong source of the nucleus enter via these two Wilson
lines.

\section{Inclusive gluon production}
From this solution, one can obtain the gluon production amplitude at
leading order as ${\mathcal M}_\lambda(\q)= q^2 {
A}^\mu(q)\epsilon_\mu^{(\lambda)}(\q)$ where
$\epsilon_\mu^{(\lambda)}(\q)$ is the polarization vector for a gluon
of 3-momentum $\q$ in the polarization state $\lambda$. Squaring,
summing over the polarizations of the produced gluon, averaging over
the sources $\rho_p$ and $\rho_{_A}$ and integrating over the impact
parameter, one obtains the following expression for the inclusive
gluon production cross-section\footnote{Our result for gluon production is
equivalent to the result of ref.~\cite{KM-DM}.}:
\begin{equation}
\omega_\q \frac{d\sigma}{d^3\q}=
\frac{\alpha_s N}{\pi^4 d_{_A} q_\perp^2}
\int\frac{d^2\k_{\perp}}{(2\pi)^2}\;
\varphi_p(\k_\perp)
\phi_{_{A}}(\q_\perp\!-\!\k_\perp)\; .
\end{equation}
In this equation, the functions $\varphi_p$ and $\phi_{_{A}}$
respectively describe the proton and the nucleus, and are expressed
as follows:
\begin{eqnarray}
&&
\varphi_p(\k_\perp)\equiv \pi^2 R_p^2 g^2 k_\perp^2 \int d^2\x_\perp
e^{i\k_\perp\cdot\x_\perp}\left<\rho_{p,a}(0)\rho_{p,a}(\x_\perp)\right>\; ,
\nonumber\\
&&
\phi_{_{A}}(\k_\perp)\equiv \frac{\pi^2 R_{_A}^2 k_\perp^2}{g^2 N}\int d^2\x_\perp
e^{i\k_\perp\cdot\x_\perp}\;
{\rm tr}
\left<U(0)U^\dagger(\x_\perp)\right>\; .
\label{eq:phis-def}
\end{eqnarray}
This formula clearly indicates that $k_\perp$-factorization remains
valid for single gluon production in collisions involving a dilute and
a dense projectile.

\section{Quark pair production in pA collisions}
One can also use the solution of the Yang-Mills equations in order to
calculate the cross-section for the production of quark-antiquark
pairs. This calculation basically amounts to finding the quark
propagator in the color field generated by the colliding projectiles
\cite{BGMP}.  For the sake of illustration, let us write here the
cross-section for single quark production\footnote{A similar formula
has been obtained in \cite{tuchin} for the special case of sources
having Gaussian correlations.} \cite{BGV2}:
\begin{eqnarray}
&&\frac{d\sigma_q}{d^2\q_\perp dy_q}=
\frac{\alpha_s^2 N}{8\pi^4 d_{_{A}}}\int\frac{dp^+}{p^+}
\int_{\k_{1\perp},\k_{2\perp}}
\frac{1}{\k_{1\perp}^2 \k_{2\perp}^2}
\nonumber\\
&&\!\!\!\!\!
\times\Big\{
{\rm tr}
\Big[(\slq\!+\!m)T_{q\bar{q}}(\k_{1\perp},\k_{2\perp})(\slp\!-\!m)
\gamma^0 T_{q\bar{q}}^{\dagger}(\k_{1\perp},\k_{2\perp})\gamma^0\Big]
\frac{C_{_{F}}}{N}\phi_{_A}^{q,q}
(\k_{2\perp})
\nonumber\\
&&\!\!\!\!\!\!\!
+\int_{\k_\perp}
\!
{\rm tr}
\Big[(\slq\!+\!m)T_{q\bar{q}}(\k_{1\perp},\k_{\perp})(\slp\!-\!m)
\gamma^0 T_{g}^{\dagger}(\k_{1\perp})\gamma^0\Big]
\phi_{_A}^{q\bar{q},g}
(\k_{2\perp};\k_\perp)+{\rm h.c.}
\nonumber\\
&&\;
+{\rm tr}
\Big[(\slq\!+\!m)T_{g}(\k_{1\perp})(\slp\!-\!m)\gamma^0 T_{g}^{\dagger}(\k_{1\perp})\gamma^0\Big]
\phi_{_A}^{g,g}(\k_{2\perp})
\Big\}
\varphi_p(\k_{1\perp})\; ,
\label{eq:cross-section-q}
\end{eqnarray}
where we use the following shorthands:
\begin{eqnarray}
&&T_{q\bar{q}}(\k_{1\perp},\k_{\perp})\equiv 
\frac{\gamma^+(\slq-\slk+m)\gamma^-(\slq-\slk-\slk_1+m)\gamma^+}
{2p^+[(\q_\perp\!-\!\k_\perp)^2+m^2]+2q^+[(\q_\perp\!-\!\k_\perp\!-\!\k_{1\perp})^2+m^2]}
\nonumber\\
&&T_{g}(\k_{1\perp})\equiv 
\frac{\slC_{_{U}}(p+q,\k_{1\perp})+\frac{1}{2}\slC_{_{V}}(p+q)}{(p+q)^2}
\; .
\label{eq:Tqqbar-Tg}
\end{eqnarray}
The distribution $\varphi_p$ that appears in this formula is the same
as the one given in eq.~(\ref{eq:phis-def}) and the function
$\phi_{_A}^{g,g}$ is identical to the function $\phi_{_A}$ of
eq.~(\ref{eq:phis-def}). However, this function is not the only piece
of information we need about the nucleus in order to compute quark
production. It turns out that we need also two other functions,
denoted here $\phi_{_A}^{q,q}$ and $\phi_{_A}^{q\bar{q},g}$. The first
one, $\phi_{_A}^{q,q}$, has a definition identical to that of
$\phi_{_A}^{g,g}$, except that the Wilson line must be evaluated in
the fundamental representation. On the contrary,
$\phi_{_A}^{q\bar{q},g}$ is defined in terms of a three-point
correlator:
\begin{equation*}
\frac{\phi_{_A}^{q\bar{q},g}(\l_\perp;\k_\perp)}{\pi R_{_A}^2}
\!=\!
\frac{2\pi l_\perp^2}{g^2 N}
\!\!\!\!\!\!\int\limits_{\x_\perp,\y_\perp}\!\!\!\!\!
e^{i(\k_\perp\!\cdot\x_\perp\!+(\l_\perp\!-\k_\perp)\!\cdot\y_\perp)}
{\rm tr}\left<\!
{\wt U}(\x_\perp)t^a {\wt U}^\dagger(\y_\perp)t^b
U_{ab}(0)
\!\right>\; ,
\end{equation*}
where ${\wt U}$ is a Wilson line in the fundamental representation
while $U$ is the adjoint Wilson line already
encountered. Eq.~(\ref{eq:cross-section-q}) is a generalization to the
case of a dense target of a well formula for proton-proton
collisions\cite{PP}. Clearly, the need for three different functions in
order to describe the nucleus in this process implies that
$k_\perp$-factorization is broken. In particular, we see that it is
not correct to simply replace the distribution function that describes
the nucleus by some saturation inspired ansatz in the formula derived
in \cite{PP}.

\section{Conclusions}
In this talk, we have seen that for proton-nucleus collisions
described in the color glass condensate framework, physical
cross-sections can be expressed in terms of correlators of Wilson
lines. It appeared that $k_\perp$-factorization holds for gluon production,
while it is broken for quark production.  In order to go further than
this analytical work and attack the case of nucleus-nucleus
collisions, one will have to perform a numerical analysis of the
Dirac equation in an external color field \cite{AA}.

\bibliographystyle{unsrt}

\end{document}